\documentstyle[12pt]{article}
\begin{document}
\input epsf 
\newcommand{\Od}{{\cal O}}
\newcommand{\lsim}   {\mathrel{\mathop{\kern 0pt \rlap
  {\raise.2ex\hbox{$<$}}}
  \lower.9ex\hbox{\kern-.190em $\sim$}}}
\newcommand{\gsim}   {\mathrel{\mathop{\kern 0pt \rlap
  {\raise.2ex\hbox{$>$}}}
  \lower.9ex\hbox{\kern-.190em $\sim$}}}
\begin{center}
\Large{\bf Dark energy in motion$^*$} \\
\vspace*{1cm} 
\large{\bf  Antonio L. Maroto}  \\
\vspace{0.3cm}
\normalsize 
Departamento de F\'{\i}sica Te\'orica\\
Universidad Complutense de Madrid\\
28040 Madrid, Spain

\vspace*{1cm}  
{\bf ABSTRACT}\\  \end{center}
Recent large-scale peculiar velocity surveys suggest that large matter
volumes could be moving with appreciable velocity with respect to the CMB rest 
frame. If confirmed, such results could conflict with the
 Cosmological Principle according to which the matter and 
CMB rest frames
should converge on very large scales. In this work we explore 
the possibility
that such large scale bulk flows are due, not to the motion of matter 
with respect to the CMB, but to the 
flow of dark energy with respect 
to matter. Indeed, when dark energy is moving,   
the usual definition of the CMB rest frame as that in which 
the CMB dipole vanishes is not
appropriate. We find instead that the dipole vanishes for observers
 at rest with respect to the {\it cosmic center of mass}, i.e. in 
motion  with respect to the background radiation.


\vspace*{0.5cm}
\noindent 
\rule[.1in]{4.5cm}{.002in}

\noindent $^*$ Essay selected for ``Honorable Mention" in the 2006 
Awards for Essays on Gravitation (Gravity Research Foundation)

\newpage
\baselineskip 20pt

In Standard Cosmology, the universe is assumed to be 
homogeneous and isotropic on very large scales. In smaller volumes, 
the presence of density inhomogeneities  generates deviations 
in the motion of 
galaxies  with respect to the pure Hubble flow. The amplitude of 
such  peculiar velocities  is determined by the amplitude
of the density perturbations. However, as we take larger and larger 
averaging volumes, we expect, according
to the Cosmological Principle, the density contrasts to decline 
and consequently  the velocity fields 
to  converge towards the pure Hubble flow. In other words, at large
scales,  
the matter rest
frame should converge to the  frame in which the expansion is 
isotropic, i.e. to the frame in which 
the CMB dipole anisotropy vanishes \cite{dipole}.

However as shown in Fig. 1, this theoretical framework is not
conclusively confirmed by  observations. Indeed, in recent years several
peculiar velocity surveys \cite{surveys} have tried to determine the volume size at which 
the streaming motion of matter with respect to the CMB vanishes. 
In  the figure the results
of different observations are compared with the {\it rms} 
expected bulk velocity
$V_b$ for  standard $\Lambda$CDM model in a sphere of radius $R$.
The results seem to agree with the theoretical expectations only at
scales $R\lsim 60h^{-1}$ Mpc. At larger scales, $R\gsim 100h^{-1}$ Mpc, 
different data sets lead to different bulk velocities both in amplitude
and direction. Moreover, there
are indeed measurements in which  large matter volumes are moving at 
speeds $\gsim 600$ km s$^{-1}$ with respect to the CMB frame, several
standard deviations away from the theoretical predictions. These results  
have been argued to be affected by systematic errors in distance
indicators, but if confirmed
by future surveys, a revision of some of the underlying ideas 
in  Standard
Cosmology would be required in order to understand the origin of
 such large flows.

\begin{figure}[h]
\centerline{\epsfxsize=8cm\epsfbox{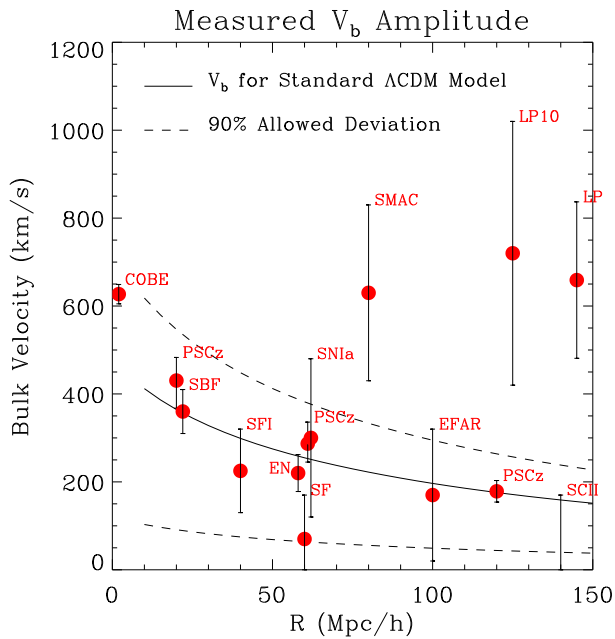}}
\caption{\footnotesize Bulk flow measurements from different peculiar
velocity surveys \cite{surveys}. The solid line corresponds to the expected {\it rms}
velocity in the standard $\Lambda$CDM cosmology, together with the
90\% deviation in dashed lines. Figure from S. Zaroubi \cite{zaroubi}}
\end{figure}

On the other hand, the main contribution to the 
CMB dipole anisotropy is usually attributed to a Doppler effect 
induced by the motion of the 
observer with respect to the last scattering surface \cite{dipole}. 
For that reason,  
bulk flow velocities are understood as the average velocity of a given 
matter volume with respect to an observer who measures a
vanishing CMB dipole. Accordingly, a different origin of the CMB dipole 
would have important consequences for the determination of cosmic flows.

So far we have only been  concerned with the possibility that 
 matter and radiation could have different rest frames even when
averaged over large volumes. However matter and radiation are
not the dominant components of the universe today. We know from 
high-redshift supernovae and other cosmological observations, that
the universe is dominated at present by some sort of dark energy
with negative pressure and whose nature is still a mistery. 
Several models have been proposed for dark energy such 
as a pure cosmological constant, quintessence or k-essence scalar
fields. 
In all those cases, dark energy
is completely decoupled from matter and radiation, its only 
effects being of gravitational  nature.

The existence of a common rest frame is expected for strongly coupled fluids,
 as is indeed
the case for baryonic matter and radiation before recombination. 
However, this might not be true at the present epoch when matter,
radiation, and presumably dark energy are almost completely 
decoupled. 
In such a case, it makes sense to explore the 
possibility that the different components have different rest frames.
In the following we will show that in such a case, the usual interpretation
of the CMB dipole is not appropriate and that even an observer at
rest with respect to the CMB could observe a non-vanishing
dipole, provided dark energy is moving with respect to matter.
This opens the interesting possibility of having non-vanishing bulk
flow velocities at large scales even if matter and radiation share
a common rest frame.

Let us therefore consider  a cosmological scenario with three perfect fluids: 
radiation, matter and dark energy, whose equations of state read 
$p_\alpha=w_\alpha \rho_\alpha$ with  $\alpha=R,M,\Lambda$. For the sake
of generality, we will allow the dark energy equation of state to
have a smooth dependence on redshift $w_\Lambda(z)$. 
The energy-momentum tensor of each fluid will take the form: 
\begin{eqnarray}
(T^\mu_{\;\;\nu})_\alpha=(\rho_\alpha+p_\alpha)
 u^\mu_\alpha u_{\nu\alpha}- p_\alpha\delta^\mu_{\;\;\nu}
\end{eqnarray}
Since  we are only interested in the effects of fluids
motion on the CMB dipole, it is sufficient to take into account
 the evolution of the largest-scale velocity
perturbations, i.e. we will just consider the zero-mode equations. The 
presence of inhomogeneities will contribute to higher multipoles. 
Therefore, for this particular problem we can write:
\begin{eqnarray}
\rho_\alpha&=&\rho_\alpha(\eta),\nonumber \\ 
 p_\alpha&=&p_\alpha(\eta), \nonumber \\ 
 u^\mu_\alpha&=&\frac{1}{a}(1,v^i_\alpha(\eta))
\end{eqnarray}    

We will assume that
$\vec v_\alpha^{\,2}\ll 1$ and we will work at first order 
in perturbation theory. In the particular case
we are considering, the form of the space-time 
metric will be given
by the following vector-perturbed spatially-flat 
Friedmann-Robertson-Walker metric:
\begin{eqnarray}
ds^2=a^2(\eta)\left(d\eta^2+2S_i(\eta)\,d\eta\, dx^i-\delta_{ij}
\,dx^i\,dx^j\right)
\label{metric}
\end{eqnarray}
 Accordingly,  
the total energy-momentum tensor reads:
\begin{eqnarray}
T^0_{\;\;0}&=&\sum_\alpha \rho_\alpha\nonumber \\
T^0_{\;\;i}&=&\sum_\alpha (\rho_\alpha+p_\alpha)(S_i-v_{i\alpha})\nonumber \\
T^i_{\;\;0}&=&\sum_\alpha (\rho_\alpha+p_\alpha)v^i_\alpha\nonumber \\
T^i_{\;\;j}&=&-\sum_\alpha p_\alpha\delta^i_{\;\;j}
\label{T}
\end{eqnarray}  
Notice that we are considering only the epoch after matter-radiation 
decoupling, assuming that dark energy is also decoupled and for that
reason we will ignore possible energy and momentum transfer
effects. 

We now calculate the linearized Einstein equations using
(\ref{metric}) and (\ref{T}).  They 
yield just the condition:
\begin{eqnarray}
S^i=\frac{\sum_\alpha (\rho_\alpha+p_\alpha)v^i_\alpha}
{\sum_\alpha (\rho_\alpha+p_\alpha)}
\label{S}
\end{eqnarray}

In General Relativity the combination 
$(\rho_\alpha+p_\alpha)$ 
appearing in (\ref{S}) 
plays the role of inertial mass density of the corresponding fluid, and 
accordingly $\vec S$ can be understood as the 
{\it cosmic center of mass velocity}. Notice that a pure cosmological
constant has no inertial mass density. 

 On the other hand, the energy conservation equations are trivially
satisfied, whereas from momentum conservation we see that the 
 velocity of each fluid relative to the center of mass frame scales
as: $\vert \vec S-\vec v_\alpha\vert\propto a^{3w_\alpha-1}$. 
Notice that for dark energy the scaling properties will depend 
on the particular
model under consideration \cite{motion}.

Once we know the form of the perturbed metric, we can calculate the
effect of fluids motion on photons propagating from the last scattering
surface using standard tools \cite{Gio}. 
The energy of a photon coming from direction 
$n^\mu=(1,n^i)$ with $\vec n^{\,2}=1$ as seen by an observer moving with
velocity $u^\mu=a^{-1}(1,v^i)$ is given by ${\cal E}=g_{\mu\nu}u^\mu P^\nu$, 
i.e. to first order in the perturbation:
\begin{eqnarray}
{\cal E}\simeq \frac{E}{a}\left(1+\frac{d\delta x^0}{d\eta}+\vec n\cdot
(\vec S- \vec v)\right)
\end{eqnarray}
where $E$ parametrizes the photon energy and the perturbed 
trajectory of the photon reads 
$x^\mu(\eta)=x_0^\mu(\eta)+\delta x^\mu$, with $x_0^\mu=n^\mu \eta$. 

In order to obtain $d\delta x^0/d\eta$, we solve the geodesics equations
to first order in the perturbations, and for 
the 0-component we get
${d^2\delta x^0}/{d\eta^2}=0$. 
By defining $\hat{\cal E}=a{\cal E}$, the temperature fluctuation 
 reads:
\begin{eqnarray}
\left.\frac{\delta T}{T}\right\vert_{dipole}&=&
\frac{\hat{\cal E}_0-\hat{\cal E}_{dec}}{\hat{\cal E}_{dec}}\simeq
\left.\frac{d\delta x^0}{d\eta}\right\vert^0_{dec}+\vec n\cdot
(\vec S- \vec v)\vert^0_{dec}\nonumber \\
&\simeq&\vec n\cdot
(\vec S- \vec v)\vert^0_{dec}
\end{eqnarray}
where the indices $0$, $dec$ denote the present and decoupling times
respectively.

At decoupling, the universe is matter dominated and we can
neglect the contribution to $\vec S$
from dark
energy. We also assume that the 
velocity of matter  is  the same as 
that of radiation, and accordingly 
we have $\vec S_{dec}\simeq \vec v_M^{dec}\simeq \vec v_R^{dec}$.
Here we  are assuming for simplicity that baryonic and dark matter share
a common rest frame.
On the other hand, neglecting contributions to the CMB dipole
of intrinsic density fluctuations in the last scattering
surface, we can 
take the emitter velocity to 
be $\vec v_{dec}\simeq \vec v_M^{dec}$. So that we find:
\begin{eqnarray}
\left.\frac{\delta T}{T}\right\vert_{dipole}&\simeq&
\vec n\cdot(\vec S_0- \vec v_0) \label{dipole}\\
&\simeq&\vec n \cdot \frac{
\Omega_M(\vec v_M^0- \vec v_0)
+(1+w_{\Lambda}^0)\Omega_\Lambda(\vec v_\Lambda^0- \vec v_0)}
{1+w_{\Lambda}^0\Omega_{\Lambda}}\nonumber
\end{eqnarray}
where $w_\Lambda^0=w_\Lambda(0)$ is the present value of the 
dark energy equation of state and we have
taken into account that today the contribution of radiation to
the energy density is negligible.

According to this result, the CMB dipole is due
to the relative velocity of the observer with respect to the present cosmic 
center of mass. When all the components share a common rest frame 
then the previous result reduces to the usual expression for the dipole:
$\delta T/T\vert_{dipole}\simeq 
\vec n\cdot(\vec v_R^0- \vec v_0)$. 
However in general
it is possible that an observer at rest with  radiation 
$\vec v_0=\vec v_R^0\neq\vec v_M^0 \neq \vec v_\Lambda^0$ can measure an
nonvanishing dipole according to (\ref{dipole}).

In the absence of dark energy or in the case in which it is in the 
form of a pure 
cosmological constant ($w_\Lambda=-1$), dark energy would not contribute 
to the center of mass motion. Moreover, today the radiation contribution
is negligible and accordingly the center of mass
rest frame would coincide with the matter rest frame.
This implies that  the relative motion of matter and radiation
today could not 
explain the existence of  bulk flows on the largest scales, since the 
frame in which the dipole vanishes would coincide with the matter
rest frame.  Conversely, the 
existence of non-vanishing bulk flows would require the presence 
of moving dark energy with $w_\Lambda^0\neq -1$. 

Indeed, if moving dark energy is responsible for the 
existence of cosmic bulk flows  on very large scales, 
then the amplitude and direction of such flows would 
provide a direct measurement
of the relative velocity of matter and dark energy. As commented
above, the bulk flow $\vec V_b$ 
can be
understood as the average velocity of a given matter volume with respect
to an observer  who  measures a vanishing CMB dipole, i.e. 
$\vec V_b=\vec v_M^0-\vec v_0$. Such an 
observer has a  velocity which is given, according to (\ref{dipole}), by:
\begin{eqnarray}
\vec v_0\simeq \vec v_M^0+\frac{(1+w_\Lambda^0)\Omega_\Lambda}
{1+w_\Lambda^0\Omega_\Lambda}
(\vec v_\Lambda^0-\vec v_M^0)
\end{eqnarray}
so that 
\begin{eqnarray}
\vec V_b \simeq \frac{(1+w_\Lambda^0)\Omega_\Lambda}
{1+w_\Lambda^0\Omega_\Lambda}
(\vec v_M^0-\vec v_\Lambda^0) 
\end{eqnarray}

Notice that, according to these results, 
even if matter is at rest with respect to the CMB radiation, 
$\vec v_M^0=\vec v_R^0$, 
it would be possible to have a non-vanishing flow $\vec V_b\neq 0$, provided
dark energy is moving with respect to matter. 

As commented above, matter and radiation were strongly coupled
in the past and for that reason it is difficult 
to understand the presence of relative motions on very large scales. 
However,  if the nature of dark energy
is really unrelated to the rest of components of the universe, 
then the corresponding primordial 
dark energy bulk velocity should be considered as a free cosmological 
parameter. That such a primordial dark energy flow could have survived until
present  giving rise to the effects studied in this work 
is a fascinating possibility \cite{motion}.

\newpage

 {\bf Acknowledgements:} 
 This work
 has been partially supported by the DGICYT (Spain) under the
 project numbers FPA 2004-02602 and FPA 2005-02327.

\end{document}